# Discriminating abilities of threshold-free evaluation metrics in link prediction

Tao Zhou[1]

CompleX Lab, University of Electronic Science and Technology of China, Chengdu 611731, People's Republic of China.

## ABSTRACT

Link prediction is a paradigmatic and challenging problem in network science, which attempts to uncover missing links or predict future links, based on known topology. A fundamental but still unsolved issue is how to choose proper metrics to fairly evaluate prediction algorithms. The area under the receiver operating characteristic curve (AUC) and the balanced precision (BP) are the two most popular metrics in early studies, while their effectiveness is recently under debate. At the same time, the area under the precision-recall curve (AUPR) becomes increasingly popular, especially in biological studies. Based on a toy model with tunable noise and predictability, we propose a method to measure the discriminating abilities of any given metric. We apply this method to the above three threshold-free metrics, showing that AUC and AUPR are remarkably more discriminating than BP, and AUC is slightly more discriminating than AUPR. The result suggests that it is better to simultaneously use AUC and AUPR in evaluating link prediction algorithms, at the same time, it warns us that the evaluation based only on BP may be unauthentic. This article provides a starting point towards a comprehensive picture about effectiveness of evaluation metrics for link prediction and other classification problems.

---

[1]Lead Contact, Correspondence: zhutou@ustc.edu

# INTRODUCTION

Network is a natural and powerful mathematical tool to characterize various social, biological and technological systems that consist of interacting elements, and network science is currently one of the most active interdisciplinary research domains [1,2]. Link prediction is an increasingly productive branch in network science, aiming at estimating existence likelihoods of missing links, future links or temporal links [3-8]. Link prediction is a paradigmatic and fundamental problem, because it is mathematically elegant and it can be treated as a touchstone for our understanding of a network's formation and evolution: In principle, better understanding will lead to more accurate prediction. Thus far, link prediction has already found many direct applications like the inference of missing biological interactions [9,10] and the online recommendation of friends and products [11,12], as well as indirect applications such as the evaluation of evolving models [13,14] and the reconstruction of networks [15,16].

A huge number of link prediction algorithms were proposed recently (see the above-mentioned surveys [3-8] and some selected representatives [17-31]). An accompanying issue is how to evaluate those algorithms [8]. As in most studies, only a very few networks are utilized to test the algorithms' performance, to enrich empirical networks is of course helpful [32-35]. On the other hand, it is critical to choose suitable evaluation metrics that can well distinguish algorithms with different predicting abilities [36,37]. This article contributes to the latter issue by proposing a novel method to measure and compare discriminating abilities of evaluation metrics.

Consider a simple network $G(V,E)$, where $V$ and $E$ are sets of nodes and links, the directionalities and weights of links are ignored, and multiple links and self-connections are not allowed. The task of link prediction is to find out missing links. To test an algorithm's accuracy, the observed links, $E$, is randomly divided into two parts: the training set $E^T$ is treated as known information, while the probe set $E^P$ is used for algorithm evaluation and no information in $E^P$ is allowed to be used for prediction, that is to say, links in $E^P$ are treated as missing links. Generally speaking, an algorithm will score all unknown links $U - E^T$, where $U$ is the universal set containing all $N(N-1)/2$ potential links, with $N = |V|$ being the network size. An well-performed algorithm is expected to assign higher scores to links in $E^P$ than to nonexistent links in $U - E$, namely to rank missing links in top positions. In practice, we usually take the top-$L$ unknown links with highest scores as predictions, where $L$ is a threshold parameter.

Evaluation metrics for link prediction can be roughly divided into two categories: threshold-dependent metrics and threshold-free metrics. The former metrics depend on threshold parameters, while the latter metrics are independent to those parameters. Precision and Recall are the two most widely used threshold-dependent metrics [38]. Precision is defined as the ratio of missing links predicted to the number of links predicted. That is to say, if we take the top-$L$ links as the predicted ones, among which $L_r$ links are correctly predicted, then the Precision equals $L_r/L$. Recall is defined as the ratio of missing links predicted to the total number of missing links, say $L_r/|E^P|$. Although threshold-dependent metrics are usually highly interpretable and of low computational complexity, they have two intrinsic disadvantages: (i) we generally do not have a reasonable way to determine the threshold, though there are some attempts [39,40], and (ii) in the comparison of algorithms' performance, different thresholds may result in different winners, so that the evaluation of algorithms based on one or a few thresholds is unauthentic. Therefore, this article focuses on threshold-free metrics. The area under the receiver operating curve (AUC) [41,42] is the most frequently used metric in link prediction, probably because of its high interpretability and good visualization. At the same time, criticisms on AUC are sharp and widespread [37,43]. For example, Hand argued that AUC is fundamentally incoherent because AUC uses different misclassification cost distributions for different classifiers [44], Lichtnwalter and Chawla commented that AUC will give misleadingly overhigh score to algorithms that can successfully rank many negatives in the bottom while this ability is less significant in imbalanced learning problem like link prediction [45], and Saito and Rehmsmeier pointed out that AUC is inadequate to evaluate the early retrieval performance which is critical in real applications especially for many biological scenarios [46] (similar concerns on binary classification were raised previously [47]). Another threshold-free metric is the area under the precision-recall curve (AUPR) [48], which becomes increasingly popular, especially for biological studies. A common advantage of AUC and AUPR is that they do not depend on any extrinsic parameters, while a common disadvantage in compared to above-mentioned threshold-dependent metrics is that the computation of AUC or AUPR is more time consuming. In addition to AUC and AUPR, a widely adopted way to evaluate algorithms is setting $L = |E^P|$ and then calculating Precision [3,18]. We name it as balanced precision (BP), since when $L = |E^P|$, Precision equals Recall. Although Precision and Recall are threshold-dependent, the intersection point of Precision-$L$ and Recall-$L$ curves itself is not dependent on $L$, and thus we also treat BP as a threshold-free metric. Notice that, although being treated as a threshold-free metric, BP still has the advantages and disadvantages of threshold-dependent metrics.

This article aims at estimating the discriminating abilities of evaluation metrics for link prediction. To achieve the goal, we need to propose a network model that simultaneously satisfy two conditions. Firstly, one can design an algorithm for the modeled networks with a parameter that can control the prediction accuracy in a monotonous way, so that one can further build the so-called discrimination matrix (see later) and then quantify metrics' discriminating abilities. Secondly, the predictability of the modeled networks should also be controllable, so that one can study the different situations for highly predictable networks and hard-to-predict networks. Our contributions are threefold. (i) We propose a method to measure the discriminating ability based on the discrimination matrix. To our knowledge, both the concept of discrimination matrix and the method to measure the discriminating ability are firstly reported in this article. (ii) We propose an algorithm with controllable accuracy for a network model with controllable predictability, on which the above-mentioned method can apply. (iii) We show that AUC and AUPR are remarkably more discriminating than BP, and AUC is slightly more discriminating than AUPR. Accordingly, we suggest that it is better to simultaneously use AUC and AUPR in evaluating link prediction algorithms, while the evaluation based only on BP may be unauthentic. Lastly, we will discuss how to extend our method to more complex network models, to real networks, to other evaluation metrics, and to more general classification problems.

## METRICS

For a binary classification problem, the receiver operating characteristic (ROC) curve is created by plotting the tree positive rate (y-axis) against the false positive rate (x-axis) at various thresholds. In the language of link prediction, given a threshold $L$, if there are $L_r$ links in $E^P$ (i.e., $L_r$ correct predictions) among the $L$ predicted links, the corresponding point in the ROC curve is $\left(\frac{L-L_r}{|U-E|}, \frac{L_r}{|E^P|}\right)$. One can thus obtain the ROC curve by varying $L$ from 1 to $|U-E^T|$. A more intuitive way to plot the ROC curve is as follows. We start from (0,0) and check all ordered unknown links one by one, from high existence likelihood to low existence likelihood. For each link, we plot an upward line with length $1/|E^P|$ if it belongs to $E^P$ or we plot a rightward line with length $1/|U-E|$ if it belongs to $U-E$. The area under the ROC curve (AUC) can be interpreted as the probability that a randomly chosen link in $E^P$ (i.e., missing links) is assigned a higher existence likelihood than a randomly chosen link in $U-E$ (i.e., nonexistent links). If all likelihoods are generated from an independent and identical distribution, the AUC value should be about 0.5. Therefore, the degree to which the value exceeds 0.5 indicates how better the algorithm performs than pure chance.

Based on the fact that the area under the ROC curve is equivalent to the probability that a randomly selected positive sample (a missing link) is scored higher than a randomly selected negative sample (a nonexistent link) [42], we can derive the analytical formula of AUC. Provided the estimated existence likelihoods of all unknown links, we assume the $|E^P|$ missing links are ranked at positions $r_1 < r_2 < \cdots < r_{|E^P|}$, with a smaller $r_i$ corresponding to a higher existence likelihood. We can introduce very small perturbations to the estimated likelihoods if there are possible deuces, so that all likelihoods are different. To calculate the probability that a missing link has a higher likelihoods than a nonexistent link, we consider all the comparisons between missing and nonexistent links. Each missing link will compare with all $|U-E|$ nonexistent links. If a missing link has rank $r_i$, there are $r_i - i$ nonexistent links ranked ahead of this missing link

(there are $r_i - 1$ links ranked ahead of this missing link, but we should remove the $i - 1$ missing links therein), namely this missing link lose $r_i - i$ comparisons. Therefore, the winning rate of this missing link is $1 - \frac{r_i-i}{|U-E|}$. Since each missing link participates in the same number of comparisons, the overall probability can be directly obtained by taking the averaging winning rate over all missing links, leading to the formula for AUC value as

$$AUC = \frac{1}{|E^P|}\sum_{i=1}^{|E^P|}\left(1 - \frac{r_i-i}{|U-E|}\right) = 1 - \frac{\langle r \rangle}{|U-E|} + \frac{|E^P|+1}{2|U-E|}, \quad (1)$$

where $\langle r \rangle = \left(r_1 + r_2 + \cdots + r_{|E^P|}\right)/|E^P|$. For example, if there are 10 unobserved links, which are ranked by estimated likelihoods as +-++--+---, where + denotes a missing link (i.e., a link in $E^P$) and - denotes a nonexistent link, namely $|E^P| = 4$, $|U - E| = 6, r_1 = 1, r_2 = 3, r_3 = 4, r_4 = 7$. According to Eq. (1), $\langle r \rangle = 15/4$ and $AUC = 19/24 \approx 0.792$. As mentioned above, AUC can be interpreted as the probability that a randomly chosen missing link has a higher existence likelihood than a randomly chosen nonexistent link, in the total $|E^P| \times |U - E| = 24$ comparisons, the missing links win 19 times, so the AUC value is 19/24. The two ways arrive to the same AUC value.

As most real networks are sparse [1,2,49], link prediction is a typical imbalanced learning problem with $|E^P| \ll |U - E|$. Therefore, AUC can be approximated as

$$AUC \approx 1 - \frac{\langle r \rangle}{|U-E^T|}, \quad (2)$$

being equivalent to the metric $\langle r \rangle / |U-E^T|$, called ranking score in the literature [50]. As real networks, especially the large-scale networks, are usually very sparse with the number of links scaling as $|E| \sim N$ [49], while $|U - E| \sim N^2$, so that the bias introduced in the approximation should scale as $N^{-1}$. In this article, we always adopt Eq. (1) to calculate AUC values, and if there are multiple links with same existence likelihoods, we can assign a random ranking to them or introduce very small random noises to break such degeneracy.

A precision-recall curve is a plot of Precision (y-axis) and Recall (x-axis) for different thresholds. When the threshold $L$ varies from $r_i$ to $r_{i+1} - 1$, Precision decreases while Recall remains unchanged, leading to a downward line, and when $L$ goes from $r_{i+1} - 1$ to $r_{i+1}$, both Precision and Recall increase, leading to an upper-rightward line. As a whole, a precision-recall curve exhibits a down-rightward sawtooth pattern ending at $(1, |E^P|/|U-E^T|)$. By summing up areas under all sawteech, one can get the value of AUPR as

$$AUPR = \frac{1}{2|E^P|}\left(\sum_{i=1}^{|E^P|}\frac{i}{r_i} + \sum_{i=1}^{|E^P|}\frac{i}{r_{i+1}-1}\right), \quad (3)$$

where $r_{|E^P|+1}$ is defined as $|U - E^T| + 1$ for convenience. According to Eq. (3), the above example results in $AUPR \approx 0.632$. In comparison with AUC, an obvious drawback of AUPR is that the absolute value of AUPR is less interpretable.

The balanced precision can be obtained by simply counting the number of missing links ranked not behind $|E^P|$, namely

$$BP = \max_i \{i | r_i \leq |E^P|\}. \quad (4)$$

In the above-mentioned example, as there are three missing links ranked no more than $|E^P| = 4$, the balanced precision is $BP = 3/4 = 0.75$.

# MODEL

In this article, we study a toy model where a simple network $G(V, E)$ is created by assigning every node pair $(i, j)$ $(1 \leq i < j \leq N)$ a link with independent probability $q_{ij}$. We focus on a simple case with every probability $q_{ij}$ is independently generated from a uniform distribution $\mathbb{U}(0, q_{max})$, where $q_{max}$ is a free parameter controlling the network density. If the probabilities $\{q_{ij}\}$ are all known, the perfect algorithm will assign a likelihood $s_{ij} = q_{ij}$ to any unknown link $(i,j) \in U - E^T$, essentially independent to the training set $E^T$ [51]. We assume that an algorithm $\Omega(\eta)$ with noise strength $\eta \geq 0$ will assign a likelihood $s_{ij} = q_{ij} + \varepsilon_{ij}$ to any unknown link $(i,j) \in U - E^T$, where the noise $\varepsilon_{ij}$ is independently generated from a uniform distribution $\mathbb{U}(-\eta, \eta)$. Notice that, as we only care about the ranking of $s_{ij}$, it is fine to deal with likelihoods less than zero or larger than 1.

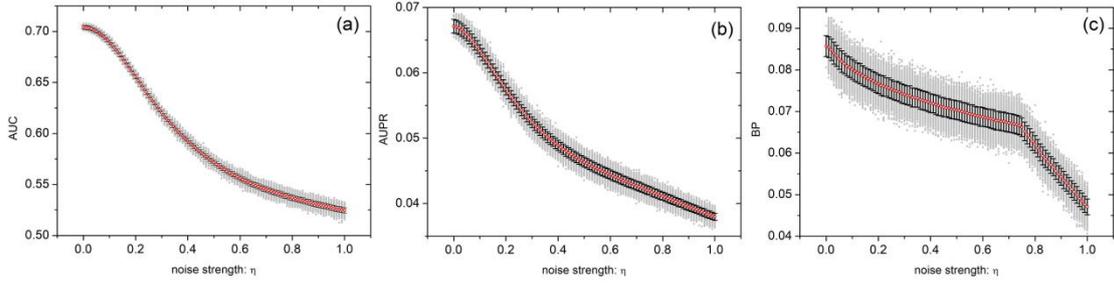

**Figure 1**. Performance of algorithms $\Omega(\eta)$ for various $\eta$, measured by (a) AUC, (b) AUPR, and (c) BP. The gray points are results for individual runs, the red circles are results averaged over 1000 runs, and the error bars show corresponding standard deviations.

In typical experiments, one run of the implementation includes the following steps: (i) generating probabilities $\{q_{ij}\}$ according to the distribution $\mathbb{U}(0, q_{max})$; (ii) constructing a network according to the generated $\{q_{ij}\}$; (iii) dividing the set of links $E$ into a training set $E^T$ and a probe set $E^P$; (iv) scoring each potential link in the set $U - E^T$; (v) computing metrics under consideration. All the experiments are implemented by a desktop with 290GHz CPU (Intel Core, two cores) and 8GB RAM. The source code and detailed annotation can be found through the URL

## RESULTS

It is obvious that the smaller the noise $\eta$ is, the better the performance of $\Omega(\eta)$ is. Figure 1 reports the performance of $\Omega(\eta)$ for various $\eta$, measured respectively by AUC, AUPR and BP. In the simulations, we set the network parameters as $N = 1000$, $q_{max} = 0.5$, and the probe ratio $\rho = \frac{|E^P|}{|E^T|+|E^P|} = 0.1$, and then we randomly generate 10 networks and for each network and each $\eta$ we implement 100 independent runs. As shown in figure 1, all the three threshold-free metrics decreases as the increase of noise strength, in consistent to our expectation. Although all the three metrics exhibit the expected tendency, their statistical features are different. The BP values obtained from the 1000 runs are most scattered, while the AUC values are most concentrated (see the grey data points in figure 1). Correspondingly, the error bars for BP are longest while those for AUC are shortest. AUPR lies in between AUC and BP (closer to AUC). In a word, AUC is slightly stabler than AUPR, and both AUC and AUPR are much stabler than BP.

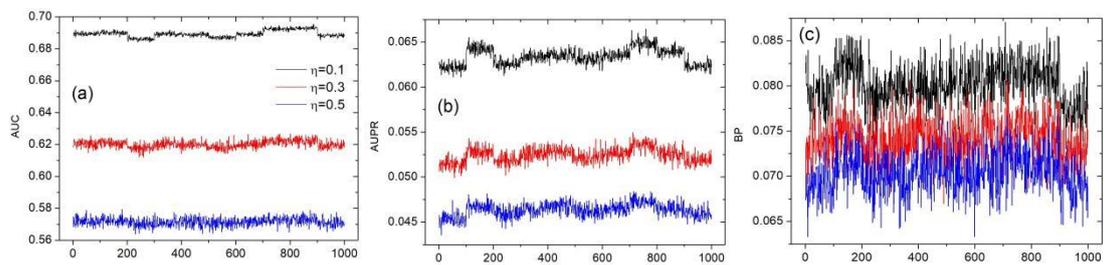

**Figure 2**. Metric values of the 1000 individual runs of (a) AUC, (b) AUPR and (c) BP. The black, red and blue curves stand for the cases of weak noise ($\eta = 0.1$), moderate noise ($\eta = 0.3$) and strong noise ($\eta = 0.5$), respectively. Other parameter settings are the same to figure 1.

Figure 2 illustrates the results of all 1000 runs for three typical noise strengths. Both AUC and AUPR can perfectly distinguish the three algorithms $\Omega(0.1)$, $\Omega(0.3)$ and $\Omega(0.5)$, while in a few runs, BP will assign higher value to $\Omega(0.3)$ than $\Omega(0.1)$, or assign higher value to $\Omega(0.5)$ than $\Omega(0.3)$. To quantitatively judge whether results of two noise strengths can be distinguished by a given metric, we apply the idea borrowed from p-value statistics. Taking a general example, if our hypothesis is $x > y$, and in 10000 independent simulations, there are 9075 times $x > y$ and 925 times $x \leq y$, then the p-value equals 0.0925. If the p-value is smaller than a given threshold $p^*$, we see the hypothesis is held at the significance level $p = p^*$. In the current issue, providing two noise strengths $\eta_1 < \eta_2$ and an evaluation metric $M$ ($M$ can be *AUC*, *AUPR*, *BP* or others), a natural expectation is $M[\Omega(\eta_1)] > M[\Omega(\eta_2)]$. If among $Z$ independent comparisons, there are $z$ times against the expectation (i.e., $M[\Omega(\eta_1)] \leq M[\Omega(\eta_2)]$), the p-value of the hypothesis that the metric $M$ can distinguish algorithms $\Omega(\eta_1)$ and $\Omega(\eta_2)$ is

$$p(\eta_1, \eta_2) = z/Z. \qquad (5)$$

Notice that, the definition of p-value is symmetric, say $p(\eta_1, \eta_2) = p(\eta_2, \eta_1)$. Analogous to a routine operation in statistics, we introduce a threshold $p^*$ and say algorithms $\Omega(\eta_1)$ and $\Omega(\eta_2)$ are distinguishable by $M$ at the significance level $p^*$ if $p(\eta_1, \eta_2) < p^*$.

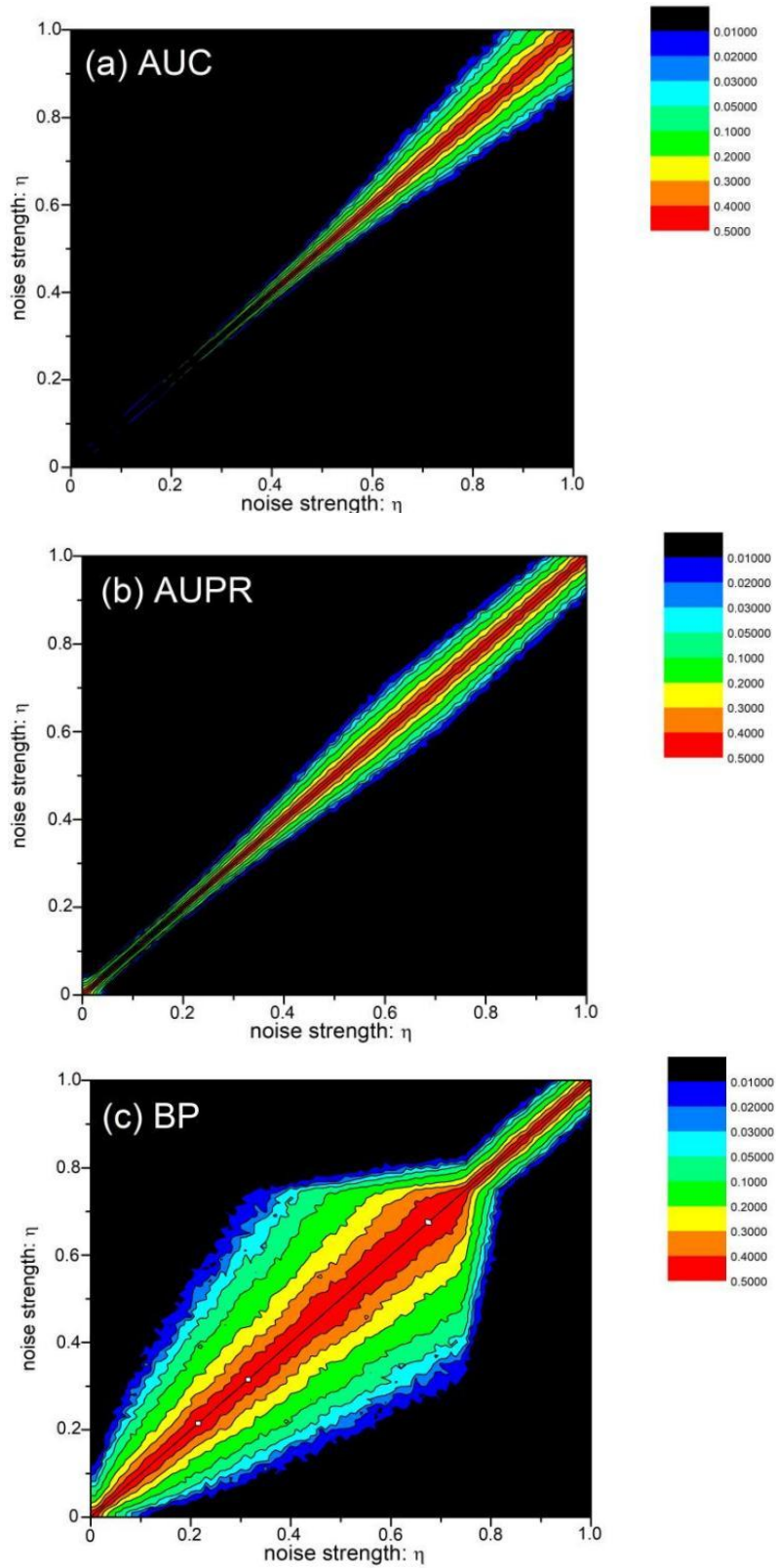

**Figure 3**. The discrimination matrices for (a) AUC, (b) AUPR and (c) BP, obtained by 1000 runs with parameter settings same to figure 1.

Given network parameters and an evaluation metric $M$, we can obtain the so-called discrimination matrix $P = \{p_{ij}\} = p(\eta_i, \eta_j)$ via Eq. (5), which defines $M$'s overall discriminating ability. Figure 3 compares the discrimination matrices of AUC, AUPR and BP. For every matrix, elements around the diagonal are usually of larger values, because if a certain element $p_{ij}$ is nearby the diagonal, then $\eta_i$ should be close to $\eta_j$ and thus $\Omega(\eta_i)$ and $\Omega(\eta_j)$ are hard to be distinguished. Direct observation shows that in a large area the p-values of BP are much higher than those of AUC and AUPR. Taking a typical threshold value $p^* = 0.01$, the black areas in the heat maps are where $(\eta_i, \eta_j)$ can be distinguished at $p^*$ level. Obviously, the black areas of AUC or AUPR is much larger than that of BP, and the black area of AUC is slightly larger than that of AUPR. Only in the area of extremely strong noise (i.e., the top right corner), the three metrics are of similar discriminating abilities.

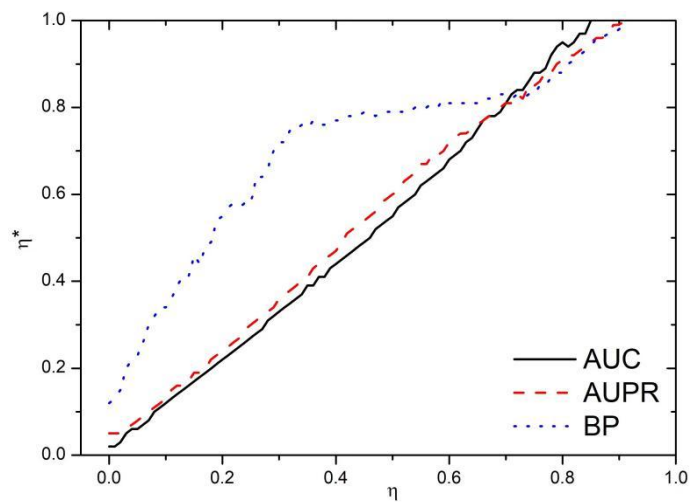

**Figure 4**. Discriminating limits for various $\eta$. The black solid line, red dash line and blue dotted line denote the results for AUC, AUPR and BP, respectively. Other parameter settings are the same to figure 1.

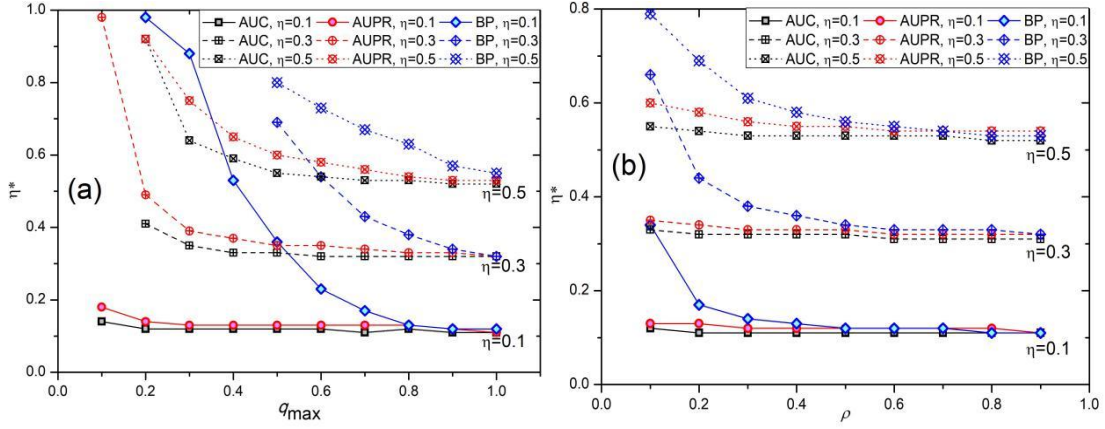

**Figure 5**. How discriminating limits $\eta^*$ change with (a) $q_{max}$ and (b) $\rho$. The solid, dash and dotted lines denote the results for $\eta = 0.1$, $\eta = 0.3$ and $\eta = 0.5$, respectively. The black, red and blue lines stand for the results for AUC, AUPR and BP, respectively. In the simulations, we set $N = 1000$ and $p^* = 0.01$, and in (a), we set $\rho = 0.1$, in (b), we set $q_{max} = 0.5$. Given $\eta$, our simulations to determine $\eta^*$ only cover the range $\eta^* \in (\eta, 1]$ and ignore the case with extremely strong noise strength $\eta^* > 1$. All results are averaged over 1000 independent runs.

Given $p^*$ and $M$, for any $\eta$, we define its discriminating limit $\eta^* > \eta$ as the minimal noise strength such that for any $\eta' \geq \eta^*$, $\Omega(\eta)$ and $\Omega(\eta')$ are distinguishable by $M$ at the significance level $p^*$. Clearly, a smaller $\eta^*$ corresponds to a higher discriminating ability. In a similar way, we can define the discriminating limit smaller than $\eta$, however, benefiting from the symmetric nature of the discrimination matrix, considering one direction $\eta^* > \eta$ is enough. As shown in figure 4, in the range $\eta \in [0,0.7]$, AUC has slightly higher discriminating ability than AUPR, and both AUC and AUPR are of remarkably higher discriminating abilities than BP. When $\eta$ exceeds 0.7, the results reverse, namely BP shows the highest discriminating ability, followed by AUPR, and AUC is the lowest. However, the discrepancy among the three metrics in such range is minor. Notice that, in the above simulations, the linking probabilities $\{q_{ij}\}$ are generated from the uniform distribution $\mathbb{U}(0,0.5)$, in comparison, the range $[0,0.7]$ already covers the cases from weak noise to strong noise, and the algorithms of extremely strong noises are less meaningful in practice.

Although results reported in figures 1 to 4 are obtained under a specific setting of parameters (i.e., $N = 1000$, $q_{max} = 0.5$, $\rho = 0.1$ and $p^* = 0.01$), we have tested the cases with parameters $q_{max} \in [0.1,1]$, $\rho \in [0.1,0.9]$ and $p^* \in [0.01,0.1]$, and demonstrated that if the number of missing links

$$|E^P| = \rho|E| \approx \frac{\rho N(N-1)}{2} \langle q_{ij} \rangle \approx \frac{1}{4}\rho q_{max} N^2 \tag{6}$$

is large enough, our findings are robust. Figure 5 shows how the discriminating abilities (measured by $\eta^*$) of the three metrics change with $q_{max}$ and $\rho$. As shown in figure 5a, given $\eta$, $\eta^*$ decreases as the increase of $q_{max}$, because larger $q_{max}$ corresponds to higher signal-noise ratio of the algorithm (recalling the likelihood $s_{ij} = q_{ij} + \varepsilon_{ij}$) and thus potentially higher discriminating ability. In figure 5b, we also see a decreasing tendency of $\eta^*$ (i.e., increasing discriminating ability) as the increase of $\rho$. In routine link prediction problem, enlarging the size of probe set will simultaneously bring two opposite effects, one decreases the prediction accuracy because the known topological information is reduced, while another one increases the prediction accuracy for algorithms that are sensitive to the imbalance between positives and negatives. As all linking probabilities are assumed to be independent to each other in the current toy model, the size of training set does not affect the algorithm and thus only the latter effect works. That is to say, the larger $\rho$ leads to more positives (i.e., missing links) so that the statistical fluctuation resulted from sample size is reduced and the algorithms are easier to be distinguished regardless the selected evaluation metric. Overall speaking, results shown in both figures 5a and 5b are consistent with out main findings. According to Eq. (6), the increasing of $N$ has similar effect as the increasing of $\rho$ or $q_{max}$, so we do not show the results by varying $N$. Indeed, we have checked the robustness of our findings up to $N = 10000$. However, this may be not held for real networks, where the likelihoods of links are not independent to each other and the size effects are complicated.

## CONCLUSION AND DISCUSSION

A careful reevaluation of performance metrics is critical for the future development of link prediction, and also helpful if one wants reexamine known results in link prediction. In this article, we propose a framework that estimates metrics' discriminating abilities based the so-called discrimination matrix, which is consisted of p-values from comparisons between many pairs of theoretically known winner and loser. Such a matrix of p-values characterizes the discriminating ability of a metric. Providing any two cases and a threshold of p-value $p^*$, we can easily judge whether the target metric can distinguish the two cases at the significant level $p^*$. We further utilize this framework to a theoretically minimum model with controllable accuracy for the link prediction algorithm and controllable predictability for the generated networks. We test the method on three widely used threshold-free metrics, and show that AUC and AUPR are remarkably more discriminating than BP, and AUC is slightly more discriminating than AUPR.

The above findings provide a valuable reference for the selection of evaluation metrics. Strictly speaking, algorithm $X$ is better than algorithm $Y$ only if $X$'s threshold curve completely dominates

$Y$'s curve, namely $Y$'s curve is always equal or below (at least somewhere below) $X$'s curve [52]. It is proved that a curve dominates in ROC space if and only if it dominates in precision-recall space [53]. However, such criterion is too strict as it is hard for an algorithm to win at every threshold. Therefore, we compare areas under threshold curves. According to our main findings, we suggest to use AUC and AUPR simultaneously and only if both AUC and AUPR of algorithm $X$ is significantly larger than algorithm $Y$, we can prudently infer that $X$ is better than $Y$. An additional suggestion is that we'd better not only use BP (or threshold-dependent metrics) unless some reliable domain knowledge advises us to do so. The findings and suggestion provide a preliminary response to the increasing debate on the effectiveness of those metrics.

To emphasize the importance of the main suggestion of this work, in Table 1, we summarize the usage of metrics in some highly influential works on link prediction. One can see clearly from Table 1 that AUC is the most frequently used metric, followed by BP, while AUPR is rarely applied. Among all above works, only Kitsak *et al.* [31] hold similar opinion to this paper, that is, one should test algorithms by both AUC and AUPR.

The toy model in this article is probably the mathematically minimum model to show the discriminating abilities of metrics. However, such model is far different from real networks that exhibit small-world effect [54], scale-free property [55], modular organization [21,56], hierarchical structure [19,57], homophily feature [58,59], and so on. Therefore, instead of the independent linking probabilities $\{q_{ij}\}$, future works should consider more complex settings of $\{q_{ij}\}$ to mimic real networks. For example, by introducing heterogeneous distribution of $\sum_j q_{ij}$

(using techniques similar to those in [60,61]), one can mimic the scale-free property in real networks where linking probabilities $\{q_{ij}\}$ are still uncorrelated. For another example, as many real networks are locally clustered with large clustering coefficients [54] (or say homophily in social networks [58]), we can introduce the triple correlation among $q_{ij}$, $q_{ik}$ and $q_{jk}$ [62].

**Table 1**: Metrics used in some selected representative works (marked by ×). Ref. [22] considered precision at a fixed threshold L = 100. Ref. [24] proposed a metric called area under the precision curve (AUP), where the precision curve shows how precision changes with the increasing threshold *L*. Ref. [29] indeed plotted the precision-recall curves, but it directly compares algorithms by corresponding precision-recall curves instead of comparing the AUPR values. Ref. [33] reported precision and recall at a threshold that maximizes the F measure for each network.

| Ref. | AUC | AUPR | BP | Others |
|---|---|---|---|---|
| [17] | × | | | |
| [18] | | | × | |
| [19] | × | | | |
| [20] | × | | | |
| [21] | × | | | |
| [22] | × | | | × |
| [23] | × | | | |
| [24] | | | × | × |
| [25] | | | × | |
| [26] | × | | × | |
| [27] | | | × | |
| [28] | × | | | |
| [29] | | | | × |
| [30] | | × | | |
| [31] | × | × | | |
| [33] | × | | | × |

Another effective way is to compare metrics' discriminating abilities based on real networks. Although the signal-noise ratio or link predictability cannot be preseted or controlled for real networks, we can use some smart skills to approximately achieve the goal. For example, after the division of $E^T$ and $E^P$, we can consider an algorithm $\Lambda(\phi)$ that makes use of only a fraction $\phi$ of randomly selected links in $E^T$ to predict missing links. Notice that, the other $(1-\phi)|E^T|$ links are completely abandoned, ensuring that $E^P$ and $U-E$ are unchanged. Since the positives and negatives are fixed, if $\Lambda(\phi)$ can to some extent capture the regularity of network organization, a natural expectation is that more known information will lead to better algorithm's performance. That is to say, if $\phi_1 > \phi_2$, $\Lambda(\phi_1)$ should outperforms $\Lambda(\phi_2)$. Then, using the present method, we can measure and compare different evaluation metrics' discriminating abilities. Notice that, when dealing with real networks, the response to the change of training set or noise strength may be very complex (e.g., not monotonous or monotonous but with much different sensitivities), so that the validity of the above expectation still asks for careful studies in the future. In addition, as in the real networks, $\{q_{ij}\}$ for unobserved links are unknown, the evaluation on metrics has to involve link prediction algorithms. This makes the analyses much more complicated while provides a chance to draw a more comprehensive picture about which metrics are better for a given real network (or a kind of real networks) and a given link prediction algorithm (or a class of algorithms).

In addition to the suggestion based on the toy model, readers should pay more attention to the proposed method, namely how to build a discrimination matrix and then quantify a metric's discriminating ability. To our knowledge, this is the first time to attempt to measure a matric's discriminating ability (early works, such as [63] and its following works, used similar terms but indeed dealt with different issues). Readers are strongly encouraged to apply this method to more candidate metrics like H measure [44] that uses a new weighting function based on the prior probabilities of missing and nonexistent links to ensure that different link predictors share the same misclassification costs, concentrated ROC [64] that calculates the area under the concentrated ROC curve with decreasing weights on low-ranked samples, and normalized Discounted Cumulative Gain [65] that depends on both a prespecified function describing relevancy of samples and a discount function describing significance of ranks. Though the threshold-free metrics studies in this article are all designed for binary classification including, future studies do not need to be limited to metrics for binary classification, and thus readers are also strongly encouraged to extend this method to analyze general classification problems.

**Acknowledgments**. This work was partially supported by the National Natural Science Foundation of China under Grant No. 11975071, and the Science Strength Promotion Programmer of UESTC under Grant No. Y03111023901014006.

**Author Contributions**. T.Z. Conceived and designed the project. T.Z. performed the experiment. T.Z. analyzed the data. T.Z. wrote the manuscript.

**Declaration of Interests**. The author declares no competing financial interests.

**References**

[1] Barabasi, A.-L. (2016). Network Science (Cambridge University Press).

[2] Newman, M. E. J. (2018). Networks (Oxford University Press).

[3] Lü, L., and Zhou, T. (2011). Link prediction in complex networks: A survey. Physica A 390, 1150-1170.

[4] Wang, P., Xu, B., Wu, Y., and Zhou, X. (2015) Link prediction in social networks: the state-of-the-art. Science China Information Sciences 58, 1-38.

[5] Martínez, V., Berzal, F., and Cubero, J. C. (2016). A survey of link prediction in complex networks. ACM Computing Surveys 49, 69.

[6] Kumar, A., Singh, S. S., Singh, K., and Biswas, B. (2020). Link prediction techniques, applications, and performance: A survey. Physica A 553, 124289.

[7] Divakaran, A., and Mohan, A. (2020). Temporal link prediction: A survey. New Generation Computing 38, 213-258.

[8] Zhou, T. (2021). Progresses and challenges in link prediction. iScience 24, 103217.

[9] Csermely, P., Korcsmáros, T., Kiss, H. J., London, G., and Nussinov, R. (2013). Structure and dynamics of molecular networks: a novel paradigm of drug discovery: a comprehensive review. Pharmacology and Therapeutics 138, 333-408.

[10] Ding, H., Takigawa, I., Mamitsuka, H., and Zhu, S. (2014). Similarity-based machine learning methods for predicting drug–target interactions: a brief review. Briefings in Bioinformatics 15, 734-747.

[11] Aiello, L. M., Barrat, A., Schifanella, R., Cattuto, C., Markines, B., and Menczer, F. (2012). Friendship prediction and homophily in social media. ACM Transactions on the Web 6, 9.

[12] Lü, L., Medo, M., Yeung, C.-H., Zhang, Y.-C., Zhang, Z.-K., and Zhou, T. (2012).

Recommender systems. Physics Reports 519, 1-49.

[13] Wang, W.-Q., Zhang, Q.-M., and Zhou, T. (2012). Evaluating network models: A likelihood analysis. EPL 98, 28004.

[14] Zhang, Q.-M., Xu, X.-K., Zhu, Y.-X., and Zhou T. (2015). Measuring multiple evolution mechanisms of complex networks. Scientific Reports 5, 10350.

[15] Squartini, T., Caldarelli, G., Cimini, G., Gabrielli, A., and Garlaschelli, D. (2018). Reconstruction methods for networks: the case of economic and financial systems. Physics Reports 757, 1-47.

[16] Peixoto, T. P. (2018). Reconstructing Networks with Unknown and Heterogeneous Errors. Physical Review X 8, 041011.

[17] Al Hasan, M., Chaoji, V., Salem, S., and Zaki, M. (2006). Link prediction using supervised learning. In Proceedings of SDM06: Workshop on Link Analysis, Counter-Terrorism and Security (Vol. 30), pp. 798-805.

[18] Liben-Nowell, D., and Kleinberg, J. (2007). The link-prediction problem for social networks. Journal of the American Society for Information Science and Technology 58, 1019-1031.

[19] Clauset, A., Moore, C., and Newman, M. E. J. (2008). Hierarchical structure and the prediction of missing links in networks. Nature 453, 98-101.

[20] Zhou, T., Lü, L., and Zhang, Y.-C. (2009). Predicting missing links via local information. The European Physical Journal B 71, 623-630.

[21] Guimerà, R., and Sales-Pardo, M. (2009). Missing and spurious interactions and the reconstruction of complex networks. PNAS 106, 22073-22078.

[22] Liu, W., and Lü, L. (2010). Link prediction based on local random walk. EPL 89, 58007.

[23] Menon, A. K. and Elkan, C. (2011). Link Prediction via Matrix Factorization. In Proceedings of the Joint European Conference on Machine Learning and Knowledge Discovery in Databases (Springer), pp. 437-452.

[24] Cannistraci, C. V., Alanis-Lobato, G. and Ravasi, T. (2013). From link-prediction in brain connectomes and protein interactomes to the local-community-paradigm in complex networks. Scientific Reports 3, 1613.

[25] Lü, L., Pan, L., Zhou, T., Zhang, Y.-C., and Stanley, H. E. (2015). Toward link predictability of complex networks. PNAS 112, 2325-2330.

[26] Pan, L., Zhou, T., Lü, L., and Hu, C. K. (2016). Predicting missing links and identifying spurious links via likelihood analysis. Scientific Reports 6, 22955.

[27] Pech, R., Hao, D., Pan, L., Cheng, H., and Zhou, T. (2017). Link prediction via matrix completion. EPL 117, 38002.

[28] Zhang, M., and Chen, Y. (2018). Link prediction based on graph neural networks. In Proceedings of the 32nd International Conference on Neural Information Processing Systems (ACM Press), pp. 5171-5181.

[29] Benson, A. R., Abebe, R., Schaub, M. T., Jadbabaie, A., and Kleinberg, J. (2018). Simplicial closure and higher-order link prediction. PNAS 115, E11221-E11230.

[30] Kovács, I. A., Luck, K., Spirohn, K., Wang, Y., Pollis, C., Schlabach, S., Bian, W., Kim, D. K., Kishore, N., Hao, T., Calderwood, M. A., Vidal, M., and Barabási, A.-L. (2019). Network-based prediction of protein interactions. Nature Communications 10, 1240.

[31] Kitsak, M., Voitalov, I., and Krioukov, D. (2020). Link prediction with hyperbolic geometry. Physical Review Research 2, 043113.

[32] Mara, A. C., Lijffijt, J., and Bie, T. De (2020). Benchmarking Network Embedding Models for Link Prediction: Are We Making Progress? In Proceedings of the 7th IEEE International Conference on Data Science and Advanced Analytics (IEEE Press), pp. 138-147.

[33] Ghasemian, A., Galstyan, A., Airoldi, E. M., and Clauset, A. (2020). Stacking models for nearly optimal link prediction in complex networks. PNAS 117, 23393-23400.

[34] Muscoloni, A., Michieli, U., and Cannistraci, C. V. (2020). Adaptive network automata modeling of complex networks. Preprint: 202012.0808.

[35] Zhou, T., Lee, Y.-L., and Wang, G. (2021). Experimental analyses on 2-hop-based and 3-hop-based link prediction algorithms. Physica A 564, 125532.

[36] Lichtenwalter, R. N., Lussier, J. T., and Chawla, N. V. (2010). New perspectives and methods in link prediction. In Proceedings of the 16th ACM SIGKDD International Conference on Knowledge Discovery and Data Mining (ACM Press), pp. 243-252.

[37] Yang, Y., Lichtenwalter, R. N., and Chawla, N. V. (2015). Evaluating link prediction methods. Knowledge and Information Systems 45, 751-782.

[38] Herlocker, J. L., Konstan, J. A., Terveen, L. G., and Riedl, J. T. (2004). Evaluating collaborative filtering recommender systems. ACM Transactions on Information Systems 22, 5-53.

[39] Liu, C., Berry, P. M., Dawson, T. P., Pearson, R. G. (2005). Selecting thresholds of occurrence in the prediction of species distributions. Ecography 28, 385-393.

[40] Jiménez-Valverde, A., and Lobo, J. M. (2007). Threshold criteria for conversion of probability of species presence to either–or presence–absence. Acta Oecologica 31, 361-369.

[41] Hanely, J. A., and McNeil, B. J. (1982). The meaning and use of the area under a receiver

operating characteristic (ROC) curve. Radiology 143, 29-36.

[42] Bradley, A. P. (1997). The use of the area under the ROC curve in the evaluation of machine learning algorithms. Pattern Recognition 30, 1145-1159.

[43] Lobo, J. M., Jiménez-Valverde, A., and Real, R. (2008). AUC: a misleading measure of the performance of predictive distribution models. Global Ecology and Biogeography 17, 145-151.

[44] Hand, D. J. (2009). Measuring classifier performance: a coherent alternative to the area under the ROC curve. Machine Learning 77, 103-123.

[45] Lichtenwalter, R. N., and Chawla, N. V. (2012). Link prediction: fair and effective evaluation. In Proceedings of the 2012 IEEE/ACM International Conference on Advances in Social Networks Analysis and Mining (IEEE Press) pp. 376-383.

[46] Saito, T., and Rehmsmeier, M. (2015). The precision-recall plot is more informative than the ROC plot when evaluating binary classifiers on imbalanced datasets. PLoS ONE 10, e0118432.

[47] Baker, S. G., Pinsky, P. F. (2001). A proposed design and analysis for comparing digital and analog mammography: special receiver operating characteristic methods for cancer screening. Journal of the American Statistical Association 96, 421-428.

[48] Davis, J., and Goadrich, M. (2006) The relationship between Precision-Recall and ROC curves. In Proceedings of the 23rd International Conference on Machine Learning (ACM Press), pp. 233-240.

[49] Del Genio, C. I., Gross, T., and Bassler, K. E. (2011). All scale-free networks are sparse. Physical Review Letters 107, 178701.

[50] Zhou, T., Ren, J., Medo, M., and Zhang, Y. C. (2007). Bipartite network projection and personal recommendation. Physical Review E 76, 046115.

[51] Garcia-Perez, G., Aliakbarisani, R., Ghasemi, A. and Serrano, M. A. (2020). Precision as a measure of predictability of missing links in real networks. Physical Review E 101, 052318.

[52] Provost, F., Fawcett, T. and Kohavi, R. (1998). The case against accuracy estimation for comparing induction algorithms. In Proceedings of the 15th International Conference on Machine Learning (Morgan Kaufmann Publishers), pp. 445-453.

[53] Davis, J., and Goadrich, M. (2006) The relationship between Precision-Recall and ROC curves. In Proceedings of the 23rd International Conference on Machine Learning (ACM Press), pp. 233-240.

[54] Watts, D. J., and Strogatz, S. H. (1998). Collective dynamics of 'small-world' networks, Nature 393, 440-442.

[55] Barabási, A. L., and Albert, R. (1999). Emergence of scaling in random network networks. Science 286, 509-512.

[56] Newman, M. E. J. (2006). Modularity and community structure in networks. PNAS 103, 8577-8582.

[57] Ravasz, E., and Barabási, A. L. (2003). Hierarchical organization in complex networks. Physical Review E 67, 026112.

[58] McPherson, M., Smith-Lovin, L., and Cook, J. M. (2001) Birds of a Feather: Homophily in Social Networks. Annual Review of Sociology 27, 415-444.

[59] Leicht, E. A., Holme, P., and Newman, M. E. J. (2006). Vertex similarity in networks. Physical Review E 73, 026120.

[60] Molloy, M., and Reed, B. (1995). A critical point for random graphs with a given degree sequence. Random Structures & Algorithms, 6, 161-180.

[61] Catanzaro, M., Boguná, M., and Pastor-Satorras, R. (2005). Generation of uncorrelated random scale-free networks. Physical Review E, 71, 027103.

[62] Newman, M. E. J. (2009). Random graphs with clustering. Physical Review Letters 103, 058701.

[63] Ling, C. X., Huang, J., and Zhang, H. (2003). AUC: a statistically consistent and more discriminating measure than accuracy. In Proceedings of the 18th International Joint Conference on Artificial Intelligence (IJCAI Press), pp. 519-524.

[64] Swamidass, S. J., Azencott, C. A., Daily, K., and Baldi, P. A. (2010). CROC stronger than ROC: measuring, visualizing and optimizing early retrieval. Bioinformatics 26, 1348-1356.

[65] Wang, Y, Wang, L., Li, Y., He, D., Chen, W., and Liu, T. Y. (2013). A theoretical analysis of NDCG ranking measures. In Proceedings of the 26th Annual Conference on Learning Theory (COLT Press), pp. 25-54.